\documentclass[numreferences]{kluwer}    

\newdisplay{guess}{Conjecture}

\usepackage{psfig}
\usepackage{klucite}

\newcommand\tfx{{F_{\rm x}}}

\newcommand\cfr{{\cal F}_{\rm r}}
\newcommand\gx{{GX~339$-$4}}
\newcommand\cyg{{Cyg~X-1}}
\newcommand\isis{{\tt ISIS}}
\newcommand\xspec{{\tt XSPEC}}
\newcommand\atca{\textsl{ATCA}}

\newcommand\rxte{\textsl{RXTE}}
\newcommand\pca{\textsl{PCA}}
\newcommand\asm{\textsl{ASM}}
\newcommand\hexte{\textsl{HEXTE}}

\newcommand\aproxgt{\mathrel{%
      \rlap{\raise 0.511ex \hbox{$>$}}{\lower 0.511ex \hbox{$\sim$}}}}
\newcommand\aproxlt{\mathrel{%
      \rlap{\raise 0.511ex \hbox{$<$}}{\lower 0.511ex \hbox{$\sim$}}}}

\begin{document}                                                                                   
\begin{article}
\begin{opening}         
\title{X-ray and Radio Monitoring of GX~339$-$4 and Cyg~X-1}
\author{Michael \surname{Nowak}}  
\runningauthor{Michael Nowak}
\runningtitle{Monitoring of GX~339$-$4 and Cyg~X-1}
\institute{Massachusetts Institute of Technology - Chandra X-ray Science Center}

\begin{abstract}
Previous work by Motch et al. (1985) suggested that in the low/hard
state of \gx, the soft X-ray power-law extrapolated backward in energy
agrees with the IR flux level.  Corbel and Fender (2002) later showed
that the typical hard state radio power-law extrapolated forward in
energy meets the backward extrapolated X-ray power-law at an IR
spectral break, which was explicitly observed twice in \gx.  This has
been cited as further evidence that jet synchrotron radiation might
make a significant contribution to the observed X-rays in the hard
state.  We explore this hypothesis with a series of simultaneous
radio/X-ray hard state observations of \gx.  We fit these spectra with
a simple, but remarkably successful, doubly broken power-law model
that indeed requires a spectral break in the IR.  For most of these
observations, the break position as a function of X-ray flux agrees
with the jet model predictions.  We then examine the radio flux/X-ray
flux correlation in \cyg\ through the use of 15 GHz radio data,
obtained with the Ryle radio telescope, and \textsl{Rossi X-ray Timing
Explorer} data, from the \textsl{All Sky Monitor} and pointed
observations.  We find evidence of `parallel tracks' in the
radio/X-ray correlation which are associated with `failed transitions'
to, or the beginning of a transition to, the soft state.  We also find
that for \cyg\ the radio flux is more fundamentally correlated with
the hard, rather than the soft, X-ray flux.
\end{abstract}

\end{opening}           

\section{Introduction}  
    
Both \cyg\ and \gx\ in their spectrally hard, radio-loud states have
served as canonical examples of the so-called `low state' (or `hard
state') of galactic black hole candidates
\cite{pottschmidt:02a,nowak:02a}.  In this state the X-ray spectrum is
reasonably well-approximated by a power-law with photon spectral index
of $\Gamma \approx 1.7$, with the power-law being exponentially cutoff
at high energies ($\approx 100$\,keV).  Such spectra have been
attributed to Comptonization of soft photons from an accretion disk by
a hot corona; however, it recently has been hypothesized that the
X-ray spectra of hard state sources might instead be due to
synchrotron and synchrotron self-Compton (SSC) radiation from a mildly
relativistic jet \cite{markoff:01a,markoff:03a}.  Jet models have been
prompted in part by multi-wavelength (radio, optical, X-ray)
observations of hard state systems.

In hard states of \gx, the 3--9\,keV X-ray flux (in units of
$10^{-10}~{\rm erg~cm^{-2}~s^{-1}}$) is related to the 8.6\,GHz radio
flux (in mJy) by $\tfx \approx 0.46 \cfr^{1.42}$ \cite{corbel:03a}.
This correlation was seen to hold over several decades in X-ray flux,
and also to hold for two hard state epochs that were separated by a
prolonged, intervening soft state outburst.  It further has been
suggested that the $\tfx \propto \cfr^{1.4}$ correlation is a
universal property of the low/hard state of black hole binaries
\cite{gallo:03a}.  This specific power-law dependence of the radio
flux upon the X-ray flux naturally arises in synchrotron jet models
\cite{falcke:95a,corbel:03a,markoff:03a,heinz:03a}, where the
optically thin synchrotron spectrum, occurring above an IR spectral
break, is presumed to continue all the way through the X-ray.

Interestingly, nearly 20 years ago Motch et al. (1985)
\cite{motch:85a} noted that for a set of simultaneous IR, optical, and
X-ray observations of the \gx\ hard state, the extrapolation of the
X-ray power-law to low energy agreed with the overall flux level of
the optical/IR data.  Corbel and Fender (2002) \cite{corbel:02a}
reanalyzed these observations (which did not include simultaneous
radio data), as well as a set of (not strictly simultaneous)
radio/IR/X-ray observations from the 1997 \gx\ hard state.  They
showed that the low energy extrapolation of the X-ray power-laws, and
the high energy extrapolation of the radio power-law, coincided with a
spectral break in the IR.

\section{Observations of \gx}\label{sec:ircoin}

We consider a set of ten simultaneous radio/X-ray observations of \gx,
eight of which come from the 1997 or 1999 hard state
\cite{wilms:99aa,nowak:02a} and and two of which come from the 2002
hard state \cite{homan:04a}.  All X-ray observations were performed
with the \textsl{Rossi X-ray Timing Explorer} (\rxte).  Note that five
of these observations are further labeled A--E, as we single these out
for special discussion. A and B occurred immediately after the 1999
soft-to-hard state transition \cite{nowak:02a} and have optically thin
radio spectra ($\alpha_{\rm r} < 0$).  C has a very `inverted' radio
spectrum (see below). D has only a single radio point, and hence we
cannot extrapolate its radio power-law without making further
assumptions.  E has the brightest X-ray flux in our sample, and is one
of the brightest hard X-ray states observed in \gx\ to date.

To analyze the X-ray spectra of these observations, \rxte\ response
matrices were created using the software tools available in {\tt
HEASOFT 5.3}, which we find yield extremely good agreement between
the \textsl{Proportional Counter Array} (\pca) and \textsl{High Energy
X-ray Timing Explorer} (\hexte) when fitting power-law models to the
Crab pulsar plus nebula system.  This is true for both the power-law
normalization and slope, both of which must be determined very
accurately when extrapolating over large energy ranges.

The radio data for observation E were obtained with the
\textsl{Australia Telescope Compact Array} (\atca) at 4.8\,GHz and
8.6\,GHz.  The radio data for observation D were also obtained with
\atca, but only at 5 GHz.  All other radio data can be found in
\cite{nowak:02a}.  

\section{A Rant on the Nature of Evil}

The observations were analyzed with the {\tt Interactive
Spectral In-} {\tt terpretation System} (\isis) \cite{houck:00a}.
For our purposes, there are several major reasons for our use of
\isis.  Data input without a response matrix (i.e., the radio data)
are automatically presumed to have an associated diagonal response
with one cm$^2$ effective area and one second integration time.  We
convert the radio data from mJy to photon rate in narrow bands around
the observation frequencies, and use this as input for the
simultaneous radio/X-ray fits. 

\begin{figure}
\hbox{\psfig{figure=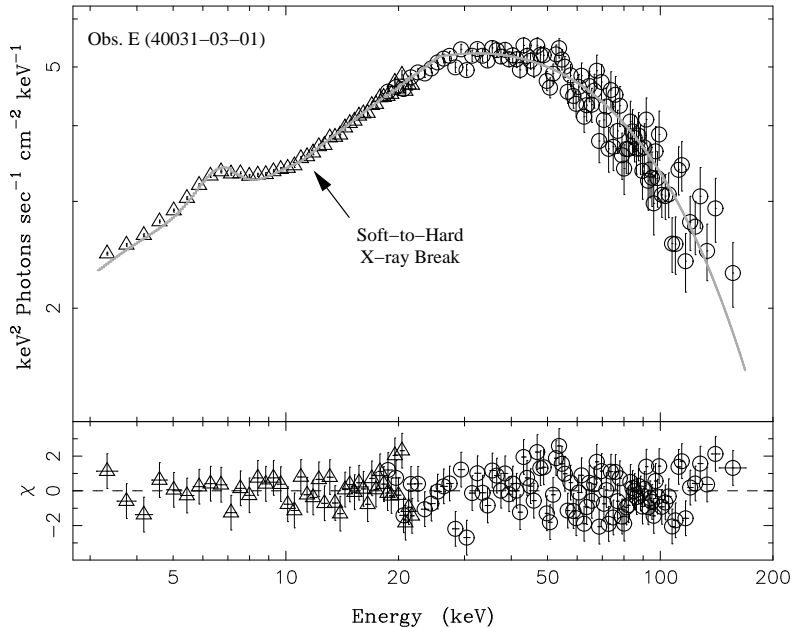,width=0.88\textwidth}}
\caption{Unfolded spectra of an X-ray spectrum of \gx\ fit with an
absorbed, exponentially cutoff, broken power-law and a gaussian
line. Residuals are from the proper forward folded model
fit.}\label{fig:unfold}
\end{figure}

The other major reason for using \isis\ is that it treats `unfolded
spectra' (shown in Fig.~\ref{fig:unfold}) in a model-independent
manner.  The unfolded spectrum in an energy bin denoted by $h$ is
defined by: $F_{\rm unfold}(h) = ( [ C(h) - B(h) ] / \Delta t )/(\int
R(h,E) A(E) dE )$, where $C(h)$ is the total detected counts, $B(h)$
is the background counts, $\Delta t$ is the integrated observation
time, $R(h,E)$ is the unit normalized response matrix describing the
probability that a photon of energy $E$ is detected in bin $h$, and
$A(E)$ is the detector effective area at energy $E$. Contrary to
unfolded spectra produced by {\tt XSPEC}, this definition produces a
spectrum that is independent of the fitted model.  Any unfolded
spectrum should be considered something of a sin; however, \isis\
unfolded spectra are only venial sins, whereas \xspec\ unfolded
spectra should rightly be classified as cardinal sins ({\sl dictum vel
factum vel concupitum contra legem aeternam}).  In
Fig.~\ref{fig:unfold}, however, the plotted residuals are those
obtained from a proper forward-folded fit.

\section{Radio-to-X-ray Break Energy Correlations}\label{sec:break}

We obtain surprisingly good fits for nine of the ten radio/X-ray
spectra using the following simple model (using the \isis/{\tt XSPEC}
model definitions): absorption (the {\tt phabs} model, with $N_{\rm
H}$ fixed to $6\times10^{21}~{\rm cm}^2$) and a high energy,
exponential cutoff (the {\tt highecut} model) multiplying a doubly
broken power-law (the {\tt bkn2pow} model, with the first break being
in the far IR to optical regime, and the second break being
constrained to the 9--12\,keV regime) plus a gaussian line (with
energy fixed at 6.4\,keV).  When considering just the X-ray spectra, a
singly broken power-law fits all ten spectra, with better results than
any of the Comptonization models that we have tried.  

\begin{figure}
\hbox{
\psfig{figure=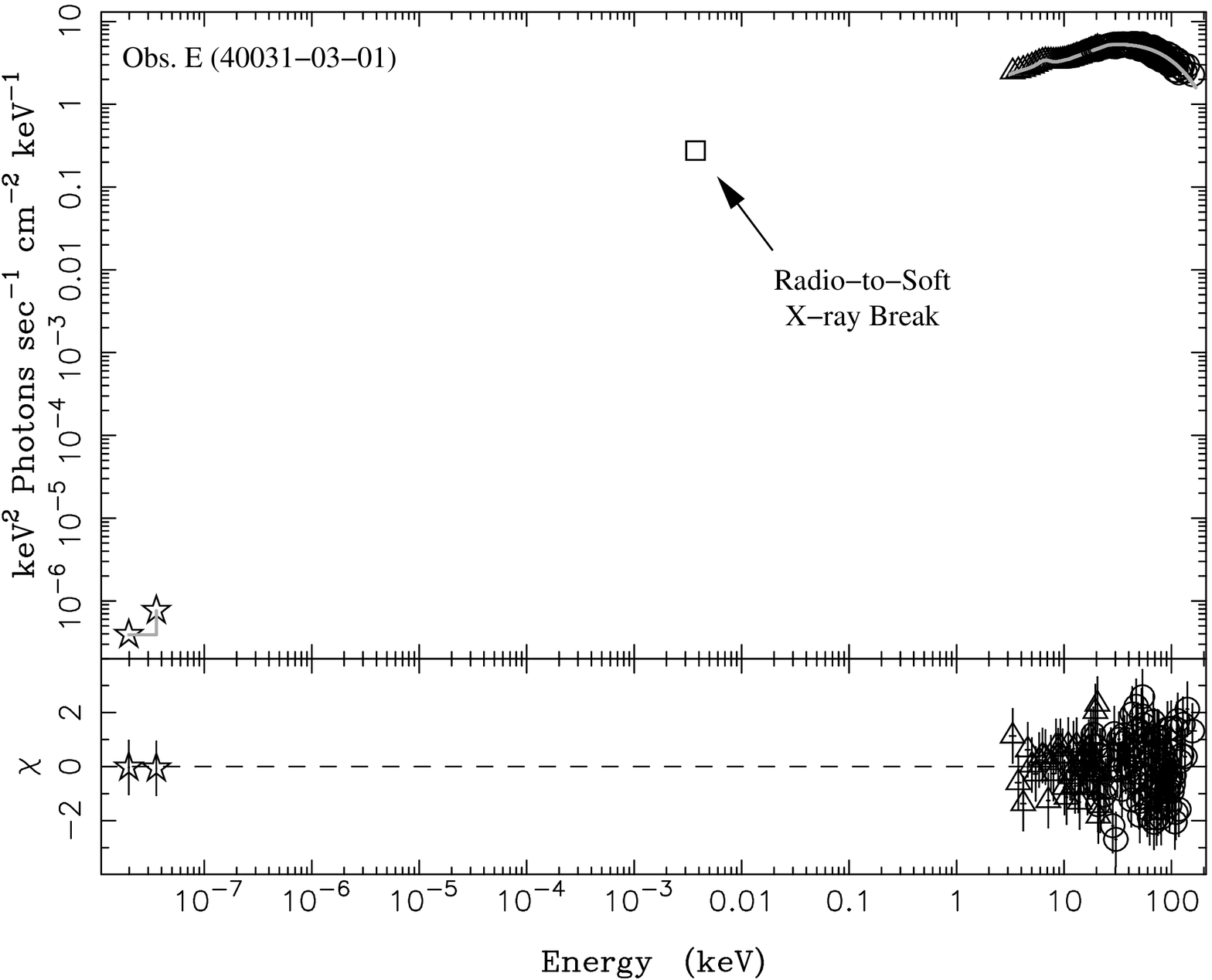,width=0.505\textwidth}
\psfig{figure=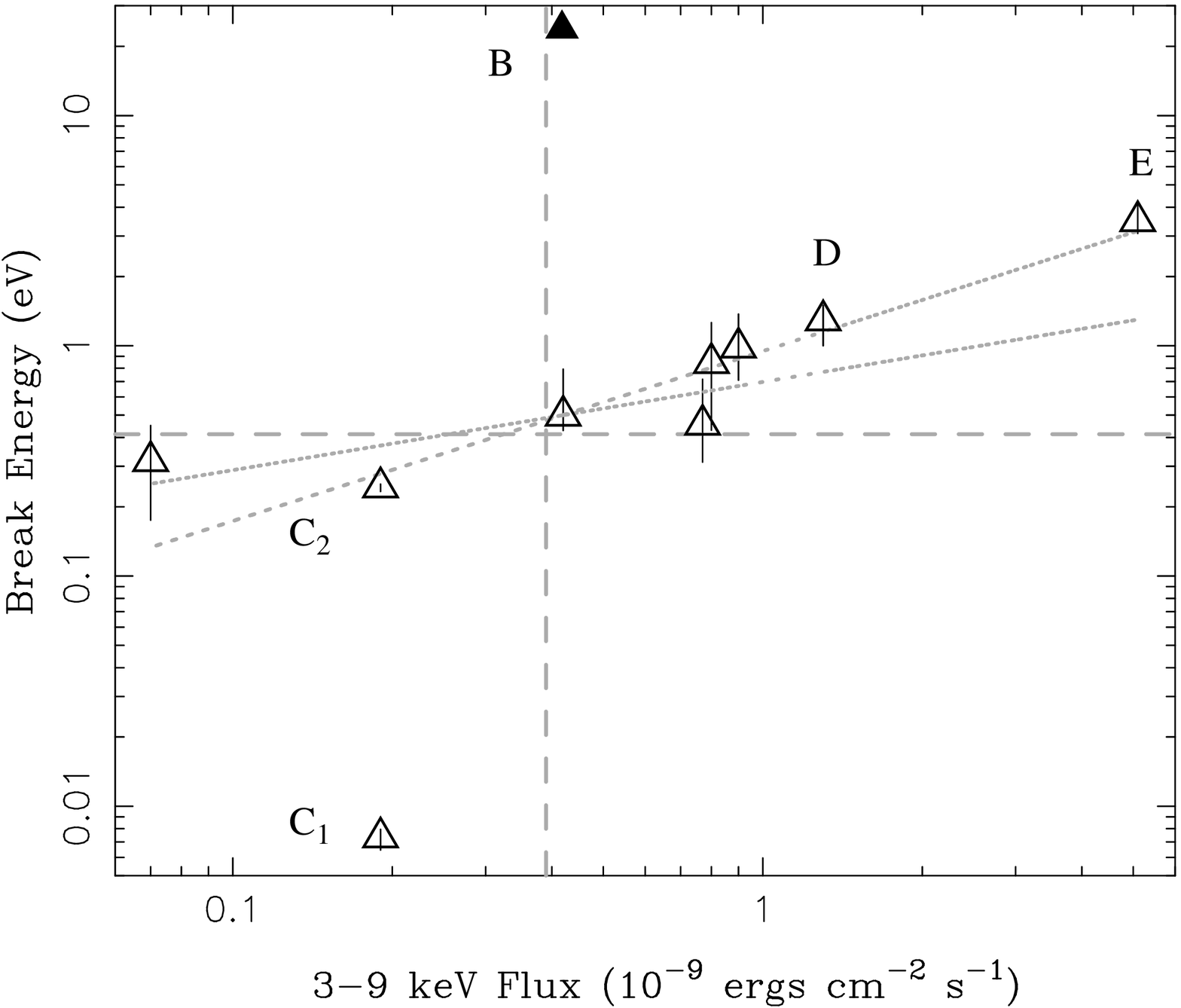,width=0.475\textwidth}
}
\caption{Left: An unfolded, simultaneous radio/X-ray observation of
\gx, fit with an absorbed, exponentially cutoff, doubly broken
power-law and a gaussian line.  Right: Results of broken power-law
fits to \gx, showing the location of the break between the radio and
soft X-ray power-law as a function of X-ray flux. Dashed lines show
the approximate integrated X-ray flux and approximate IR spectral
break energy previously observed in \gx\ (Corbel \& Fender
2002). Dotted lines are ${\rm E_{b-r}} \propto \tfx^{0.38}$, ${\rm
E_{b-r}} \propto \tfx^{0.74}$.}\label{fig:corr}
\end{figure}

In Fig.~\ref{fig:corr} we show the fitted radio-to-X-ray break
location as a function of 3--9\,keV integrated flux.  We also show in
this figure the approximate integrated 3--9\,keV flux and the IR break
location for the 1997 observation discussed by Corbel and Fender
(2002) \cite{corbel:02a}.  For our \gx\ observations of comparable
3--9\,keV flux, the doubly broken power-law models do indeed produce a
break in the IR.  The model fits presented here have predicted
radio-to-X-ray breaks ranging all the way from the far IR to the blue
end of the optical (and into the X-ray, if one also considers
observation B, which has an `optically thin' radio spectrum).

The data point labeled ${\rm C_1}$, with a break in the far IR, has an
extremely `inverted' radio spectrum ($\alpha_{\rm r} = 0.58$). This
drives the fitted break to low energies, and hence leads to deviations
from the overall observed trends shown in Fig.~\ref{fig:corr}.  Such
an inverted spectrum is very unlikely to be intrinsic to the radio
jet, and is most likely a signature of free-free absorption at low
frequencies \cite{fender:01a}.  If we instead consider only the
highest observed radio frequency (8.6\,GHz, which is likely less
affected by free-free absorption), and fix the radio spectral slope at
this point to $\alpha_{\rm r} = 0.1$, similar to the other
observations, we obtain an IR break frequency (labeled ${\rm C_2}$)
that is consistent with the other inferred breaks.

To assess the correlation of radio-to-X-ray break energy with
integrated X-ray flux, we exclude the data points from observation C
(likely free-free absorbed), observation B (which has an optically
thin radio spectrum), and observation D (which is consistent with the
trends if we assume a radio slope of $\alpha_{\rm r} = 0.1$).  A
regression fit to the remaining six data points suggests that the
radio-to-X-ray break energy, in eV, scales with the 3--9\,keV
integrated flux as $0.95 \tfx^{0.74\pm0.05}$.

Using the scale invariance {\sl Ansatz} to describe the jet physics
\cite{heinz:03a,heinz:04a}, we show elsewhere \cite{nowak:05a} that
the predicted scaling between the integrated X-ray synchrotron flux
and the radio-to-X-ray break frequency where the jet becomes optically
thin to synchrotron self-absorption scales as $\nu_{\rm b} \propto
\tfx^{2(p+6)/(p+4)/(p+5)}$, where $p$ is the power-law index of the
electron spectrum, and we have used for the X-ray spectral slope
$\alpha_{\rm x}=(1-p)/2$ from standard synchrotron theory.  For the
usual range of $-0.65 < \alpha_{\rm x} < -0.5$ of synchrotron spectra,
we obtain $\nu_{\rm b} \propto F_{\rm x}^{0.36}$ to $\nu_{\rm b}
\propto F_{\rm x}^{0.38}$.  This prediction is flatter than the
observed dependence of extrapolated break frequency upon X-ray flux.
However, if one also excludes the highest flux point, then the scaling
becomes more consistent with the jet synchrotron prediction, i.e.,
$\nu_{\rm b} \propto \tfx^{0.38\pm0.16}$ (Fig.~\ref{fig:corr}).

\section{Radio/X-ray Correlations in Cyg~X-1}\label{sec:cygcor}

\begin{figure}
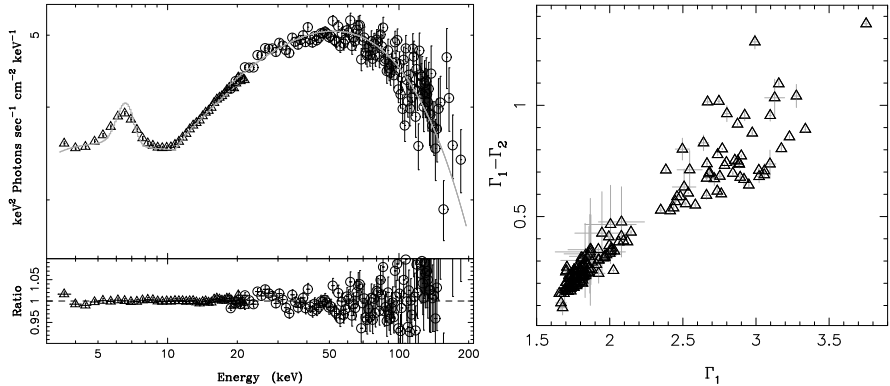

\hbox{
\psfig{figure=mnowak_fig3a.ps,width=0.53\textwidth}
\psfig{figure=mnowak_fig3b.ps,width=0.45\textwidth}
}
\caption{Left: Unfolded spectra of an X-ray spectrum of \cyg\ fit with
an absorbed, exponentially cutoff, broken power-law and a gaussian
line. Right: Correlation of the soft X-ray spectral slope with the
hard minus soft X-ray slope for all our \cyg\
observations.}\label{fig:unfold_cyg}
\end{figure}

We now turn to radio/X-ray observations of \cyg\
\cite{pottschmidt:02a,gallo:03a}.  Again, these spectra are remarkably
well-fit by a simple, exponentially cutoff broken power-law
(Fig.~\ref{fig:unfold_cyg}).  The degree of the break here indicates
that the broken power-law models are \emph{not} simply mimicking
reflection, but are suggesting two separate continuum components.  In
fact, the correlation between the soft X-ray slope and the hard minus
soft X-ray slope is more pronounced than any correlation between
reflection fraction and spectral slope (Fig.~\ref{fig:unfold_cyg}).  We
shall elaborate upon these point further in an upcoming paper (Wilms
et al., in prep.).

The associated radio data are 15\,GHz observations performed at the
Ryle Telescope, Cambridge (UK) \cite{pottschmidt:02a}.  (These are
single channel observations, so a radio spectral slope cannot be
determined.)  Most of these observations have occurred simultaneously
with pointed \rxte\ observations \cite{pottschmidt:02a}, and nearly
all have very good contemporaneous coverage by the \rxte\ \textsl{All
Sky Monitor} (\asm).


\begin{figure}
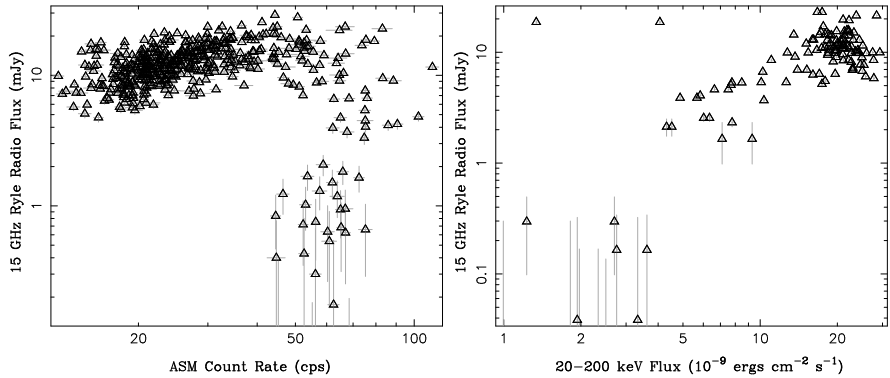

\hbox{
\psfig{figure=mnowak_fig4a.ps,width=0.49\textwidth}
\psfig{figure=mnowak_fig4b.ps,width=0.49\textwidth}
}
\caption{Left: 15 GHz Ryle radio flux (mJy) vs. \cyg\ daily mean \asm\
count rate. Right: 20--200\,keV flux (units of $10^{-9}~{\rm
ergs~cm~s^{-1}}$) vs. the daily average 15\,GHz Ryle radio flux (mJy)
for pointed observations of \cyg.}\label{fig:cyg}
\end{figure}

In Fig.~\ref{fig:cyg} we plot the daily average \asm\ count rate
vs. the daily average 15\,GHz flux.  Ranging from approximately
10--50\,cps in the \asm\, there is a clear log-linear correlation
between the radio flux and the \asm\ count rate.  As for \gx, the
radio flux rises more slowly than the \asm\ count rate ($\cfr$ scales
approximately as the 0.8 power of the \asm\ count rate). \cyg,
however, shows much more scatter in the amplitude of the correlation
than does \gx. 

As noted elsewhere \cite{gallo:03a}, there is a sharp roll-over for
higher \asm\ count rates.  However, one can clearly discern on the
shoulder of this roll-over (i.e., the upper right corner of
Fig.~\ref{fig:cyg}) four `spokes', consisting of 2--5 data points
each.  In these spokes, the radio/X-ray correlation appears to hold to
high count rates.  We have confirmed \cite{nowak:05a} that each of
these times are associated with `failed transitions' to the soft state
\cite{pottschmidt:02a}, except for the lowest amplitude of these
spokes, which occurs immediately preceding a prolonged soft state
outburst.

In Fig.~\ref{fig:cyg} we also plot the daily average \asm\ count rate
vs. the 20--200\,keV flux from our pointed \rxte\ observations taken
during the same 24 hour period \cite{pottschmidt:02a}.  We see that
hard X-ray/\asm\ correlation traces a similar pattern to the
radio/\asm\ correlation.  Indeed, when we plot the hard X-ray flux
vs. the daily average radio flux we obtain a log-linear relationship,
as shown in Fig.~\ref{fig:cyg}.  In \cyg, the radio flux density
appears fundamentally to be tied to the hard X-ray emission.

\section{Summary}

We have considered ten simultaneous \rxte/radio hard state
observations of \gx, and over one hundred \rxte/radio observations of
\cyg.  We have fit the former spectra with a very simple, but
remarkably successful, phenomenological model consisting of a doubly
broken power-law with an exponential roll-over plus a gaussian line.
For \gx, the break between the radio and soft X-ray power-laws occurs
in the IR to optical range, in agreement with prior work
\cite{motch:85a,corbel:02a}.  In contrast to prior works, we have fit
the X-ray data in `detector space' and provided a quantitative
assessment of the extrapolated break location.

The scaling of the radio-to-X-ray break location with integrated X-ray
flux agrees reasonably well with predictions of jet models wherein a
large fraction of the soft X-ray flux is due to synchrotron emission
from the jet.  At least some fraction of the observed soft X-rays may
be attributable to emission from the jet, as opposed to disk or
corona.  On the other hand, we have evidence in the \cyg\ failed state
transitions and soft state transition, that the correlation between
radio flux and integrated X-ray flux can take on different amplitudes
during different hard state episodes.  There is also evidence in \cyg\
that the radio/X-ray correlation is more fundamental to the hard X-ray
band.  In jet models, this band, which essentially encompasses the
third, highest energy, power-law component in our model fits (and also
encompasses the exponential cutoff), is possibly attributable to the
synchrotron self-Compton (SSC) emission from the base of the jet
\cite{markoff:03a,markoff:04a}.  It is therefore quite reasonable to
expect a strong coupling between the radio and hard X-ray flux;
however, these models are more complex than simple pure synchrotron
models, and are only now beginning to be explored quantitatively
\cite{markoff:03a,markoff:04a}.

The results presented here suggest, at the very least, some obvious
observational strategies.  Given the break energy correlations, it
would be extremely useful to have not only a radio amplitude for each
X-ray observation, but also a radio slope.  Furthermore, the predicted
break for the brightest observation of \gx, E, occurs in the blue end
of the optical.  Thus, ideally multi-wavelength observations would
consist of radio, broad band X-ray, and IR through optical coverage.
This is an admittedly difficult task, but BHC are demonstrating via
spectral correlations that all these energy regimes are fundamentally
related to activity near the central engine.

Finally, it is important to obtain multi-wavelength observations of
multiple episodes of each of the spectral states.  For example, if
there are indeed `parallel tracks' in the radio/X-ray correlations, it
would be interesting to determine whether the amplitude of the
radio/X-ray correlation is related to the flux at which the
outbursting source transits from the low/hard to high/soft state.  If
such observations can be made with more quantitative detail, we will
have vital clues to determining the relative contributions of coronae
and jets, and the coupling between these two components, for black
hole binary systems.

\acknowledgements This talk later evolved into the paper cited as
Nowak et al. (2004).  I would therefore like to thank my coauthors,
J. Wilms, S. Heinz, G. Pooley, K. Pottschmidt, and S. Corbel.  It is
also a pleasure to acknowledge useful conversations with Sera Markoff
and Jeroen Homan.  This work has been supported by NASA grants
SV3-73016 and GO4-5041X, and NSF grant INT-0233441.

\end{article}


\end{document}